\documentstyle[12pt]{article}
\def\mathfont#1{\ifmmode{#1}\else{$#1$}\fi} 
\def\lae{\mathrel{<\kern-1.0em\lower0.9ex\hbox{$\sim$}}}  
\def\gae{\mathrel{>\kern-1.0em\lower0.9ex\hbox{$\sim$}}}  

\begin{document}
\input psfig.sty

\noindent
{\bf \Large 
Heating a Distant Galaxy Cluster by  
Giant X-ray Cavities and Large-Scale Shock Fronts }

\bigskip
\bigskip

\noindent
{\bf B. R. McNamara$^1$, P. E. J. Nulsen$^{2,3}$, M. W. Wise$^4$, D. A. Rafferty$^1$, C. Carilli$^5$,
C. L. Sarazin$^6$, \& E. L. Blanton$^{6,7}$}
\bigskip

\noindent
$^1$ Astrophysical Institute and Department of Physics \& Astronomy, Ohio University, Clippinger Laboratories,
Athens, OH 45701\\

\noindent
$^2$ Harvard-Smithsonian Center for Astrophysics, 60 Garden St., Cambridge, MA \\

\noindent
$^3$ On leave from the University of Wollongong\\

\noindent
$^4$ MIT Center for Space Research\\

\noindent
$^5$ National Radio Astronomy Observatory, Very Large Array, Soccorro, NM\\

\noindent
$^6$ Astronomy Department, University of Virginia, Box 3818, Charlottesville,
VA 22903\\

\noindent
$^7$ Institute for Astrophysical Research, Boston University\\

\baselineskip=0.2in
\bigskip
\bigskip
\bigskip
\noindent
{\bf
Most of the baryons in galaxy clusters reside between the
galaxies in a hot, tenuous gas$^{1}$. The densest gas in their centers 
should cool and accrete onto giant central galaxies at rates 
of $10-1000$ solar masses per year$^{1}$.  However, no viable
repository for this gas has been found$^{1}$.
This paradigm changed abruptly when new X-ray observations 
showed far less cooling below X-ray temperatures than expected $^{2}$.
Consequently, most of the gas must be heated and maintained 
above $\simeq 2$ keV$^{3}$.
The most promising heating mechanism 
concerns powerful radio jets emanating from supermassive black 
holes in central cluster galaxies$^{4}$. Here we report the discovery of 
giant cavities and shock fronts in a distant ($z=0.22$) cluster
caused by an interaction between a radio source and the hot
gas surrounding it.  The energy involved is 
$\sim 6\times 10^{61}$ erg, the most powerful radio outburst known. 
This is enough energy to quench a cooling flow for several Gyr,
and to provide $\sim 1/3$ keV per particle of heat to the
surrounding cluster.}

Cavities with diameters ranging from a few to a few tens of kpc have
been found in the hot gas surrounding nearly two dozen
galaxies, groups, and clusters$^{5}$.  
Their enthalpy (free energy), which ranges between 
$pV=10^{55}-10^{60}~{\rm erg}$, scales in proportion to the cooling 
X-ray luminosity and the radio power of the host system$^{5}$. 
The cavities and weak shocks in half of these systems 
are currently injecting enough energy  into the hot gas to 
balance radiation losses (cooling)$^{6,7}$, but
it is unclear whether they can quench cooling over longer timescales. 
Systems without cavities today could have been heated 
in the past by powerful but relatively rare outbursts$^{8}$.  
However, prior to the discoveries of large scale cavities
and shocks discussed here and in
a companion paper$^{9}$, little evidence existed to support this conjecture.

Giant cavities, each roughly 200 kpc in diameter, were found in
a {\it Chandra} X-ray image the
optically poor cluster MS0735.6+7421 (Fig. 1).
The center of the cluster harbors
a cD galaxy that hosts a radio source roughly 550 kpc in size.
The radio lobes fill the cavities suggesting,
as in other clusters, that the gas was displaced and compressed
by the advancing radio source.  Both the cavities and the radio source
dwarf the cD galaxy, which is itself a member of the class of
the largest galaxies in the Universe.

The  average pressure surrounding the
cavities (Fig. 2) is $P\simeq 6\times 10^{-11}~{\rm erg~cm^{-3}}$.  The
work required to inflate each cavity against this pressure
is $pV\approx 10^{61}$ erg. Their enthalpy
can approach $4pV$ per cavity, depending on the equation of
state of the gas filling them, giving a value of $\approx 8\times 10^{61}$ erg.
To place this figure in perspective, it exceeds the enthalpy
of the cavities in the Perseus cluster$^{6}$, Cygnus A$^{5}$, and M87$^{7}$ 
by roughly 250 times, 15 times, and more than four orders of magnitude,
respectively.  Only the energy of the
recently-discovered shock front in Hydra A$^{9}$ falls within 
an order of magnitude of this value.

The bright elliptical region surrounding the X-ray cavities 
strongly resembles the hot cocoon of jet-powered radio source
models$^{10,11}$.
In this interpretation, the relatively sharp edge of the cocoon
lies at the location of an enveloping shock (Fig. 3).
The two red ``hot spots'' in Fig. 4 show 
that the gas near the cavities is being heated by the shocks.
By contrast, the gas surrounding the cavities in other systems such as 
Hydra A$^{4}$ and Perseus$^{6}$ is relatively
cool, suggesting that buoyancy, not excess pressure, is driving their
outward advance.

The shock properties were determined using a spherical
hydrodynamic model of a point explosion at the
center of an initially isothermal, hydrostatic atmosphere (Fig. 3).  
The age and driving energy of the shock are $t_{\rm s} \simeq 1.04\times
10^8$ yr and $E_{\rm s} \simeq 5.7\times10^{61}$ erg, respectively.  
The energy is proportional to the preshock temperature, which probably
exceeds 5 keV. The spherical model underestimates the shocked volume,
tending to underestimate total energy.  Furthermore, since the
cavities occupy a large fraction of the cocoon, they appear to be
driving the shock, which undermines to some degree our assumption
of a point explosion.  Nevertheless, we expect the energy of the
outburst to be within a factor $\sim2$ of the model estimate, and more
likely to exceed it.  

The shock energy is
reassuringly close to the cavity enthalpy, and
we adopt the shock energy as the probable value. 
The average jet power of the
outburst is then $P_{\rm s} = E_{\rm s} / t_{\rm s} \simeq 1.7\times
10^{46}\rm\
erg\ s^{-1}$, comparable to a powerful quasar radiating at
the Eddington limit of a $\sim 2\times 10^8~{\rm M}_{\odot}$ black hole.
The flux density of the radio source at 1.4 GHz
is 21 mJy$^{12}$, corresponding 
to a monochromatic luminosity of  $4\times 10^{40}~{\rm erg~s^{-1}}$,
several orders of magnitude fainter than powerful
quasars$^{13}$.  The ratio of average jet power to monochromatic
radio power is  $\sim 10^5$, 
enormously larger than the generally accepted factor of 10 to 100$^{13,14}$
for powerful radio sources.  Evidently, even relatively faint radio sources
can be mechanically powerful$^5$.
Bright radiation from an active quasar
is surprisingly absent both in the optical and X-ray bands.  
However, the cD harbors a 
$\sim 10^{42}~{\rm erg~sec}^{-1}$ optical
emission nebula extending over its inner 20 kpc$^{15}$, 
which are commonly found in cooling flows. 

This outburst released enough energy
to quench a 200 ${\rm M_{\odot}~yr^{-1}}$ cooling flow for
several Gyr, assuming all of the energy is deposited within 
the $\sim 50$ kpc cooling region of the cluster.  
The central cooling time of the gas is roughly 1 Gyr, so the
cooling that ensues could establish a feedback cycle of heating and cooling
driven by accretion onto the central black hole
$^{8,16,17}$.  A similar process occurring in other
clusters would in principle maintain the
observed levels of hot gas with short cooling times, molecular 
gas$^{18}$, and star formation$^{19}$ in cD galaxies, 
while preventing the development
of a more massive, steady cooling flow$^{1,20}$. 
The existence of bright nebular emission located in the
the cD galaxy is consistent with this picture$^{15}$.

Much of the energy is, however, leaving the cooling
region, bound for the cluster's outskirts.
The gas mass $5.5\pm 0.7 \times 10^{13}~{\rm M_{\odot}}$
within one Mpc is being heated at the level of
$\simeq 1/3$ keV per particle. This one event alone then provides
a substantial fraction of the $1-3$ keV per particle of heat
required to raise the entropy above the level of gravity
alone (preheating)$^{21}$.  For a bolometric, unabsorbed
X-ray luminosity of $1.1\times 10^{45}~{\rm erg~ s^{-1}}$ and
a mean temperature of $4.5$ keV, the cluster
departs upward in luminosity by several times 
from the relatively tight correlation between X-ray luminosity
and gas temperature$^{22,20}$.  Since the shock power exceeds 
the cluster's X-ray luminosity by $15$ times, 
the shock is surely capable of causing this departure.   The time required
to radiate away the shock energy $\gae E_{\rm s}/L_{\rm x}\sim 2$
Gyr is a substantial fraction of the age of the cluster.  
Therefore, 
this outburst will leave a persistent  mark on the temperature
and luminosity of the cluster, even after the cavities 
have disappeared.  Assuming this event is not unique,
substantial heating must
have occurred recently in clusters, not just during an early preheating
epoch$^{21}$.  Events of this nature would complicate 
the use of X-ray temperature and luminosity functions
to probe the large scale structure and cosmology$^{22}$.  

The titanic proportion of this event
suggests it was powered by accretion onto a black hole.
Equating the shock energy to $0.1 M c^2$ gives an estimate of the minimum
accreted mass required to power the burst of
$M \simeq 3 \times 10^8\rm\ M_\odot$,  itself the mass of
a supermassive black hole.  The relationship
between galactic bulge luminosity and black hole mass$^{23}$ predicts
a $\sim 10^9 \rm\ M_\odot$ black hole
resides there (assuming the cD's absolute visual magnitude
within a 35 kpc diameter is $-22.4^{24}$). 
Evidently the central  black hole accreted a substantial fraction
of its own mass in only $10^8$ yr, a remarkable growth rate for such a large 
black hole.  While a similar line of reasoning would apply to
quasars, their ages and average jet powers have not been measured
directly.  Such a rapid rate of growth may be difficult to 
reconcile with the small scatter in the black hole mass to bulge mass
relation$^{25}$, and is at variance
with the view that the most massive black holes have evolved slowly
in the recent past$^{26}$.

Finally, the magnetic field strengths in clusters are typically 
a few $\mu$G$^{27}$, and evidence is growing for the existence of 
large-scale, intergalactic fields$^{28,29}$. These
fields could be generated and dispersed by outflows from 
supermassive black holes$^{28,29}$. The equivalent
magnetic field strength corresponding to the energy density
within the cavities is $\sim 100 ~ \mu$G, considerably
larger than the accumulated field strengths in clusters$^{27}$.  
Therefore, a plausibly small fraction of the total energy
of this one radio outburst alone channeled into magnetic field would
magnetize the cluster.


\bigskip
\bigskip

\noindent
1. Fabian, A. C. Cooling Flows in Clusters of Galaxies, {\it Ann. Rev. Astron. Astrophys.} {\bf 32}, 277-318 (1994)

\noindent
2. Peterson, J. R. {\it et al.} High-Resolution X-Ray Spectroscopic Constraints on Cooling-Flow Models for Clusters of Galaxies. {\it Astrophys. J.} {\bf 590}, 207-224 (2003)

\noindent
3. Fabian, A. C., Mushotzky, R. F., Nulsen, P. E. J., Peterson, J. R. 
``On the soft X-ray spectrum of cooling flows.'' {\it Mon. Not. R. Astr.} {\bf 321}, 20 (2001)

\noindent
4. McNamara, B. R. {\it et al.} Chandra X-Ray Observations of the Hydra A Cluster: An Interaction between the Radio Source and the X-Ray-emitting Gas. {\it Astrophys. J. Lett.} {\bf 534}, 135-138 (2000) 

\noindent
5. B\^{\i}rzan, L. {\it et al.} A Systematic Study of Radio-induced X-Ray Cavities in Clusters, Groups, and Galaxies. {\it Astrophys. J.} {\bf 607}, 800-809 (2004)

\noindent
6. Fabian, A. C. {\it et al.} A deep Chandra observation of the Perseus cluster: shocks and ripples. {\it Mon. Not. R. astr.} {\bf 344}, 43-47 (2003)

\noindent
7. Forman, W. {\it et al.}  Reflections of AGN Outbursts in the Gaseous Atmosphere of M87. {\it Astrophys. J.} Submitted (2004)

\noindent
8. Soker, N.  {\it et al.} A Moderate Cluster Cooling Flow Model. {\it Astrophys. J.} {\bf 549}, 832-839 (2001)

\noindent
9. Nulsen, P. E. J.  {\it et al.} The Cluster-Scale AGN Outburst in Hydra A. {\it Astrophys. J.} Submitted (2004)

\noindent
10. Scheuer, P. A. G. Models of extragalactic radio sources with a continuous energy supply from a central object. {\it Mon. Not. R. astr.}  {\bf 166}, 513-528 (1974)

\noindent
11. Heinz, S., Reynolds, C. S., \& Begelman, M. C. X-Ray Signatures of Evolving Radio Galaxies. {\it Astrophys. J.} {\bf 501}, 126-136 (1998)

\noindent
12. Condon, J. J. {\it et al.} The NRAO VLA Sky Survey. {\it Astron. J.} {\bf 115}, 1693-1716 (1998)

\noindent
13. Bicknell, G. V., Dopita, M. A., \& O'Dea, C. P. Unification of the Radio and Optical Properties of Gigahertz Peak Spectrum and Compact Steep-Spectrum Radio Sources. {\it Astrophys. J.} {\bf 485}, 112-124 (1997)

\noindent
14. De Young, D. S. On the relation between Fanaroff-Riley types I and
II radio galaxies. {\it Astrophys. J. Lett. } {\bf 405}, 13-16 (1993)

\noindent
15. Donahue, M., Stocke, J. T., \& Gioia, I. M. Distant cooling flows. {\it Astrophys. J.} {\bf 385}, 49-60 (1992)

\noindent
16. Churazov, E. {\it et al.} Evolution of Buoyant Bubbles in M87. {\it Astrophys. J.} {\bf 554}, 261-273 (2001)

\noindent
17. Churazov, E. {\it et al.} Cooling flows as a calorimeter of active galactic nucleus mechanical power. {\it Mon. Not. R. astr.} {\bf 332}, 729-734 (2002)

\noindent
18. Edge, A. C. The detection of molecular gas in the central galaxies of cooling flow clusters. {\it Mon. Not. R. astr.} {\bf 328}, 762-782 (2001)

\noindent
19. McNamara, B. R. \& O'Connell, R. W. Star Formation in Cooling Flows in Clusters of Galaxies. {\it Astron. J.} {\bf 98}, 2018-2043 (1989)

\noindent
20. Donahue, M. \& Stocke, J. T. {\it ROSAT} Observations of Distant Clusters of Galaxies. {\it Astrophys. J.} {\bf 449}, 554-566 (1995)

\noindent
21. Wu, K. K. S., Fabian, A. C., \& Nulsen, P. E. J.  Non-gravitational heating in the hierarchical formation of X-ray clusters {\it Mon. Not. R. astr.} {\bf 318}, 889-912 (2000)

\noindent
22. Markevitch, M. The LX-T Relation and Temperature Function for Nearby Clusters Revisited {\it Astrophys. J.} {\bf 504}, 27-34 (1998)

\noindent
23. Gebhardt, K. {\it et al.} A Relationship between Nuclear Black Hole Mass and Galaxy Velocity Dispersion. {\it Astrophys. J. Lett.} {\bf 539}, 13-16 (2000)

\noindent
24. Stocke, J. T.  {\it et al.} The Einstein Observatory Extended Medium-Sensitivity Survey. II - The optical identifications. {\it Astrophys. J. Suppl.} {\bf 76}, 813-874 (1991)

\noindent
25. Ferrarese, L. \& Merritt, D. A Fundamental Relation between Supermassive Black Holes and Their Host Galaxies. {\it Astrophys. J.} {\bf 539}, L9-L12 (2000)

\noindent
26. Heckman, T. M. {\it et al.} Present-Day Growth of Black Holes and Bulges: the SDSS Perspective. {\it Astrophys. J.} in press (2004) astro-ph\/0406218

\noindent
27. Clarke, T. E., Kronberg, P. P., \& B\" ohringer, H. A New Radio-X-Ray Probe of Galaxy Cluster Magnetic Fields. {\it Astrophys. J. Lett. } {\bf 547}, 111-114 (2001)

\noindent
28. Kronberg, P. P. {\it et al.} Magnetic Energy of the Intergalactic Medium
    from Galactic Black Holes. {\it Astrophys. J.} {\bf 560}, 178-186 (2001)

\noindent
29. Furlanetto, S. R. \& Loeb, A. Intergalactic Magnetic Fields from Quasar Outflows. {\it Astrophys. J.} {\bf 556}, 619-634 (2001)

~~\\
All correspondence should be addressed to Brian R. McNamara,
Astrophysical Institute and Dept. of Physics \& Astronomy,
Ohio University, Clippinger Labs, Athens, OH 45701

~~\\
B.R.M thanks Gus Evrard, Dave De Young, and Joe Shields for helpful discussions.  
This work was supported by a NASA Long Term Space 
Astrophysics grant, a Chandra Archival Research grant,
a Chandra Guest Observer grant, and a contract 
from the Department of Energy through the Los Alamos National Laboratory.

\newpage 

\begin{figure}[h]
\hbox{
\hspace{.0in}
\psfig{figure=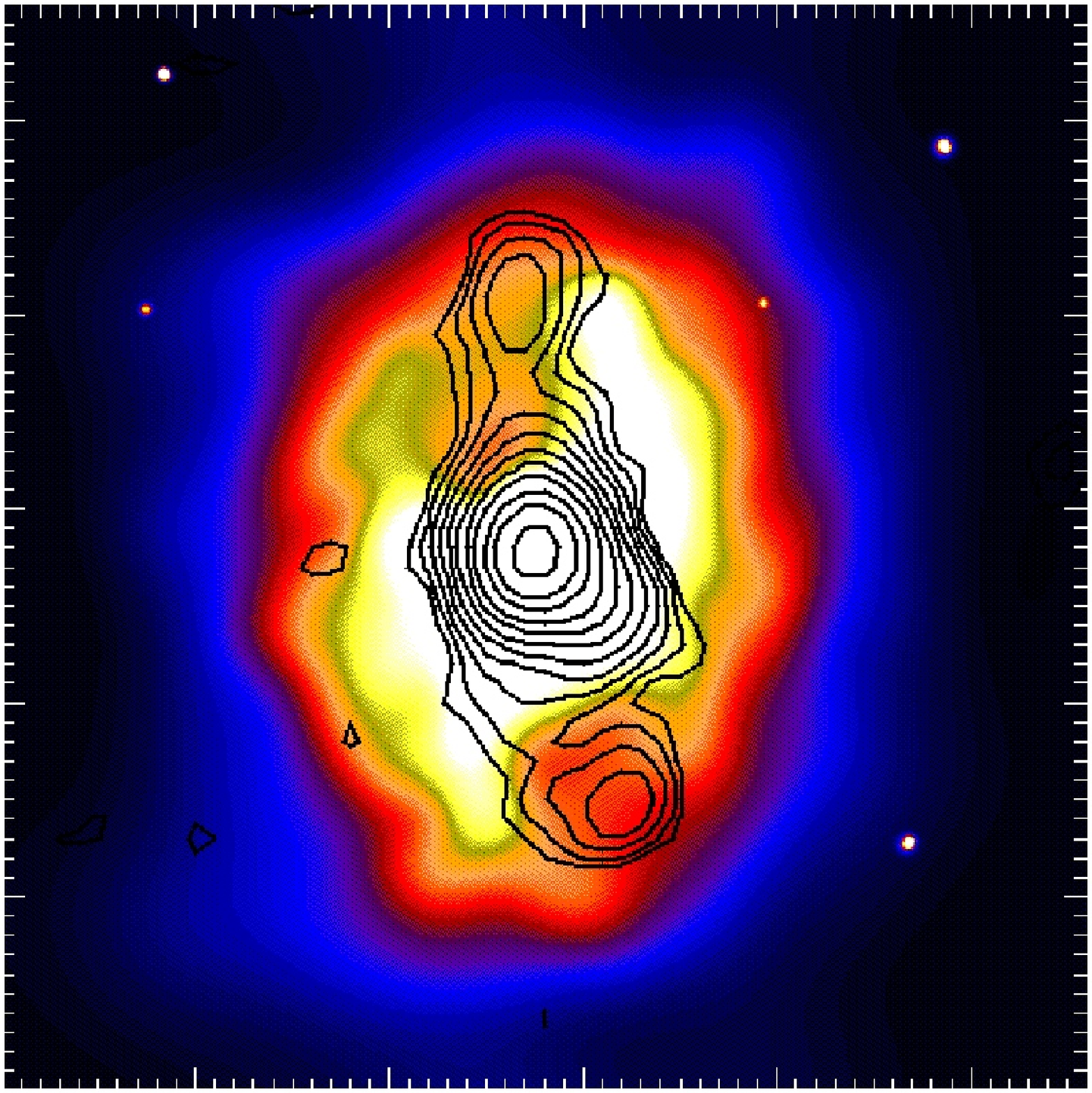,height=8truecm,width=8truecm,angle=90}
\psfig{figure=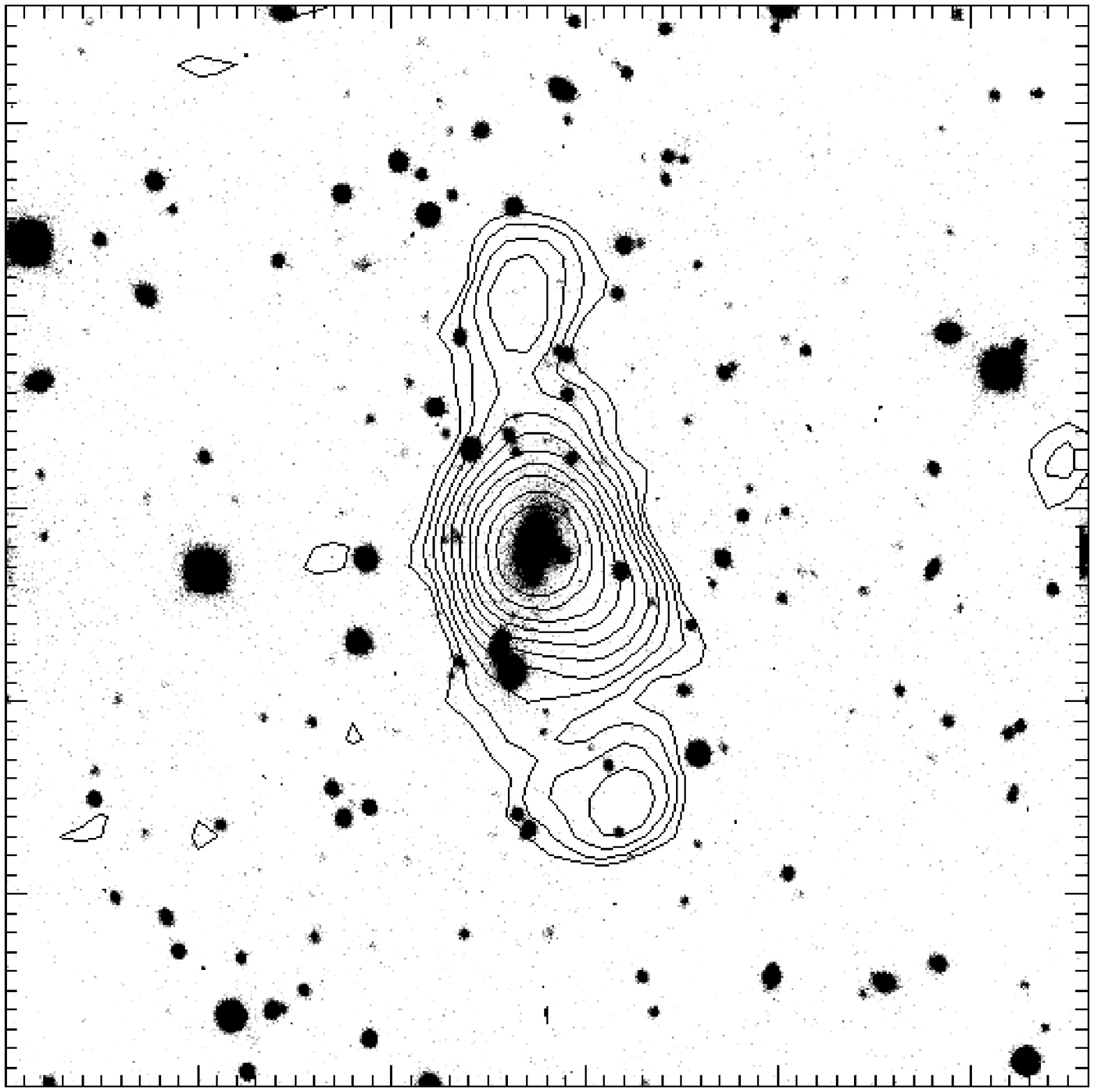,height=8truecm,width=8truecm}
}
\begin{minipage}[h]{6.0truein}
~~\\
\baselineskip=0.2in
{\bf Figure 1} The relationships between the X-ray, radio, and
optical emission of the cluster.  The smoothed X-ray image (left) and
optical image (right) are superposed with the 1.4 GHz radio contours.
The 40 ksec X-ray image was obtained with the {\it Chandra} 
X-ray Observatory on 1 December 2003.  Approximately
75000 useful X-ray photons were detected.
The X-ray surface brightness depressions (cavities)
are between $10\%-20\%$ fainter than the surrounding X-ray emission
to the north and south of the center. Most of the cluster's 
emission emerges from an elliptical structure bounded by a shock front. 
We assume a flat cosmology with $H_0=70~{\rm
  km~s^{-1}~Mpc^{-1}}$ and $\Omega=0.3$, corresponding to a ratio of
linear to angular size of 3.5 kpc arcsec$^{-1}$ at the redshift of
the cluster throughout this paper.

The   $\simeq 4$ arcsec resolution radio map 
was made with the Very Large Array telescope in the C
configuration. The
cavities are filled with radio emission. Assuming spherical cavities
whose edges lie at the midpoints of the rims,
each is roughly an arcmin in diameter (200 kpc) centered
approximately 125 kpc to the north-east and to the south-west of 
the cluster center RA = 07 41 44.0, Dec = $+$74 14 38.3, (J2000).  
The radio contour levels are $2\times 10^{-4}\times ( -1, 1,1.4,2,2.8,4,5.7,8,11,16,22,32)$ Jy/beam.  The radio contour levels on the
visual image  are $2\times 10^{-4}\times ( -1, 1,1.4,2,2.8,4,5.7,8,11,16,22)$ Jy/beam. The cD's $R$-band surface brightness 
peaks at its center at  roughly 19.5 mag arcsec$^{-2}$ and diminishes
with radius in $r^{1/4}$-law fashion, reaching 25.5  mag 
105 kpc).  Each image is $250 \times 250$ arcsec ($875\times 875$
kpc) on a side. 

\end{minipage}
\end{figure}

\newpage

\begin{figure}[h]
\hbox{
\hspace{.2in}
\psfig{figure=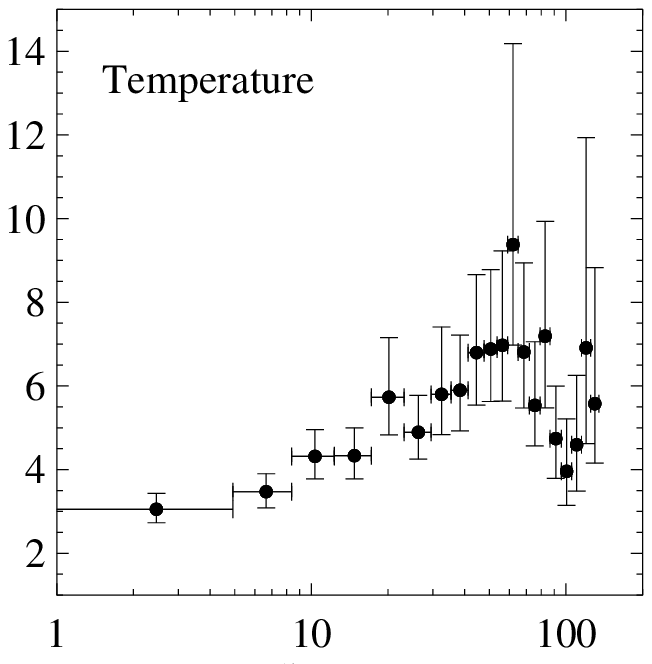,height=5truecm,width=5truecm}
\hspace{.5in}
\psfig{figure=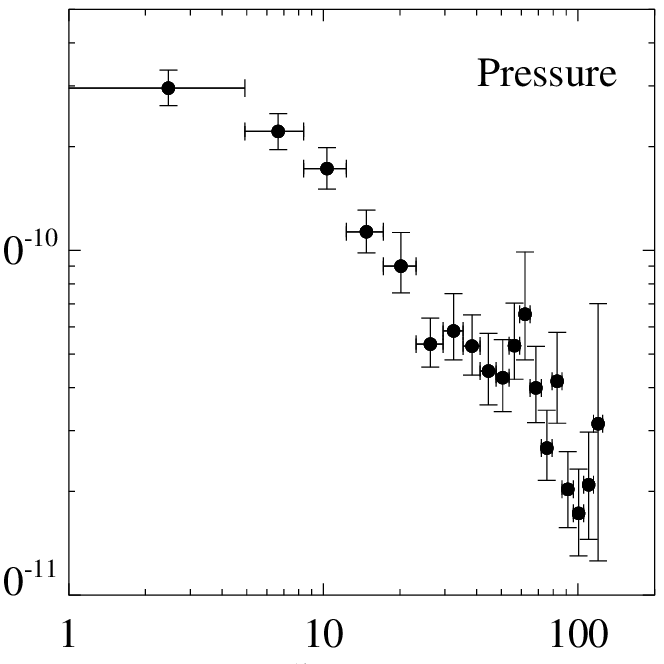,height=5truecm,width=5truecm}
}
\begin{minipage}[h]{6.0truein}
\baselineskip=0.2in
~~~\\

{\bf Figure 2} 
Projected temperature and pressure profiles of the
cluster gas. The
vertical error bars are $90\%$ confidence intervals, and the horizontal
bars represent the bin sizes. The spectra used to construct these profiles
were extracted from the X-ray image in circular apertures centered on
the cD.  The spectra were
modeled as thermal emission from a single temperature plasma of
uniform metallicity, attenuated by the foreground column of neutral hydrogen
in our Galaxy.  The average metallicity of the
gas was found to be 0.4 times the solar value. The temperature
and pressure profiles are complex.  The coolest gas located in the center
of the cluster is roughly 3 keV. The temperature rises with
increasing radius reaching an average of
about 7 keV between $50$ and $80$ arcsec.  Beyond 80 arcsec
the temperature drops to roughly 5 keV, although the level of this
drop is uncertain.  The pressure is highest in the
center $3\times 10^{-10}~{\rm erg~cm^{-3}}$, and declines smoothly
with radius in roughly power law fashion until 
reaching a radius of roughly 70 arcsec.  The pressure there 
rises abruptly at the shock front.   

\end{minipage}
\end{figure}

\newpage

\begin{figure}[h]
\hbox{
\hspace{1.2in}
\psfig{figure=fig3.ps,height=5.5truecm,width=5.5truecm,angle=-90}
}
\begin{minipage}[h]{6.0truein}
~~~\\
\baselineskip=0.2in
{\bf Figure 3} 
Projected 0.5--7.5
keV radial surface brightness profile of the cocoon region 
compared to shock model predictions. The profile was measured
in $30^\circ$ sectors to the east and west, along the minor axis of
the ellipse (PA's $90^\circ$ -- $120^\circ$ and $270^\circ$ --
$300^\circ$), where the shock is relatively uniform.
The feature at a radius of $69$
arcsec (240 kpc) is consistent with being a weak shock.  The surface
brightness beyond is well fitted by the power law, $r^{-\beta}$ with $\beta =
2.24\pm 0.29$ (90\%). To be
consistent with the surface brightness profile beyond the shock, the
gas density initially has $\rho(r) \propto r^{-\eta}$, with $\eta =
1.62$, and the gravitational field
was chosen to make the undisturbed atmosphere hydrostatic.  The model
assumes the temperature of the unshocked gas is 5 keV. The
{\it Chandra} 0.5 -- 7.5 keV response was computed using XSPEC and an
absorbed mekal model with foreground column density
$3.49\times10^{20}\rm\ cm^{-2}$, redshift 0.216 and abundance 0.4
times solar (results are insensitive to these parameters).  Model
surface brightness profiles are scaled to match the observed profile
in the unshocked region.  The three lines represent model
profiles for shock Mach numbers of 1.41 and $1.41 \pm 0.07$
(increasing Mach number from bottom to top).
The greatest source of uncertainty in the age
is the preshock temperature, since the Mach number is not strongly
model-dependent (in particular, if the cocoon is axially symmetric,
the age estimate is not sensitive to projection effects).  
The temperature increase for a shock of this magnitude
is expected to be roughly $30\%$ above the preshock value.  This jump
is consistent with the data, but the uncertainty
in the pre- and post-shock temperatures are too large to further constrain
the shock properties.  The
vertical error bars are $90\%$ confidence intervals, and the horizontal
bars represent the bin sizes.

\end{minipage}
\end{figure}

\newpage

\begin{figure}[h]
\hbox{
\hspace{.0in}
\psfig{figure=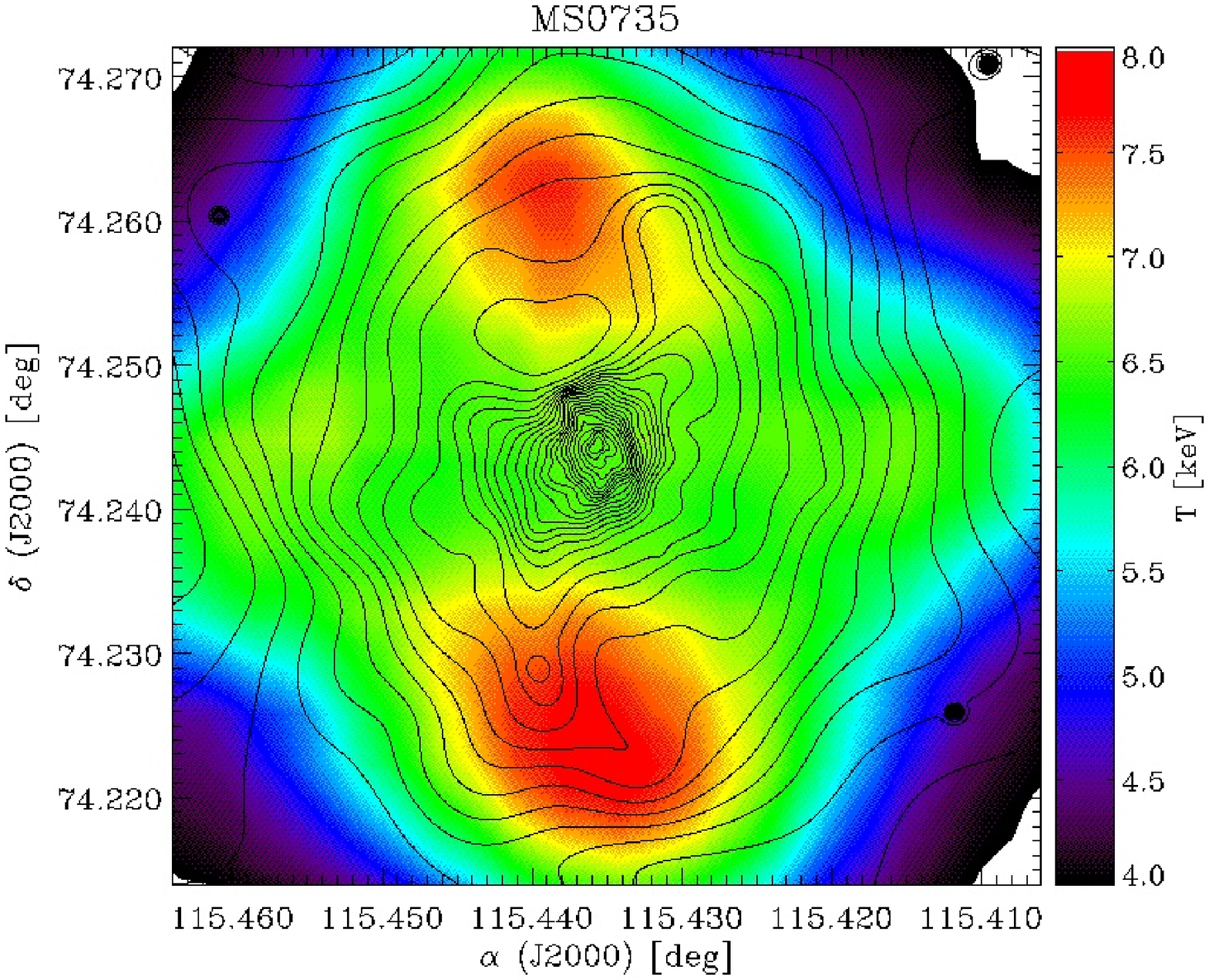,height=7.5truecm,width=9.0truecm}
}
\begin{minipage}[h]{6.0truein}
\baselineskip=0.2in
{\bf Figure 4} Temperature map of the central $200 \times 200$ 
arcsec of the cluster. Redder colors indicate
hotter temperatures.  The logarithmically spaced contours,
ranging in surface brightness 
from $1\sigma =  1.0^{-4}~ {\rm cnt ~s^{-1}~cm^{-2}~pix^{-2}}$
to $100\sigma$, show the locations of the cavities.
The hottest regions are at the tips of the cavities, where the
shocks are strongest.

\end{minipage}
\end{figure}

\end{document}